\documentclass[aps,prl,twocolumn,preprintnumbers]{revtex4-2}

\usepackage{amsmath,graphicx,amssymb}
\usepackage{xcolor} 

\def\beq{\begin{equation}}
	\def\eeq{\end{equation}}

\newcommand{\pr}[1]{ #1}

\def\deltabar{{\mathchar '26\mkern -10mu\delta}}

\begin{document}
	
	\title{Dark Matter Minihalos from Primordial Magnetic Fields}
	
	\author{Pranjal Ralegankar}
	\email{pralegan@sissa.it}
	\affiliation{SISSA, International School for Advanced Studies,
		via Bonomea 265, 34136 Trieste, Italy}
	
	\begin{abstract}
		Primordial magnetic fields (PMF) can enhance baryon perturbations on scales below the photon mean free path. However, a magnetically driven baryon fluid becomes turbulent near recombination, thereby damping out baryon perturbations below the turbulence scale. In this Letter, we show that the \pr{initial growth in baryon perturbations gravitationally induces growth in the dark matter perturbations}, which are unaffected by turbulence and eventually collapse to form $10^{-11}-10^3\ M_{\odot}$ dark matter minihalos. If the magnetic fields purportedly detected in the blazar observations are PMFs generated after inflation and have a Batchelor spectrum, then such PMFs \pr{could potentially} produce \pr{dark matter} minihalos.
	\end{abstract}
	
	\maketitle

	\noindent {\bf Introduction.}
	Magnetic fields are found to be ubiquitous in our observations of the Universe. An intriguing possibility is that the observed cosmic magnetic fields are seeded by magnetic fields produced in the early Universe, either during inflation or phase transition \cite{Vachaspati:2020blt,Subramanian:2015lua}. So far no direct evidence for the primordial nature of the magnetic fields has been obtained. However, the absence of GeV gamma-ray halos around TeV blazars might be evidence for primordial magnetic fields \cite{doi:10.1126/science.1184192,HESS:2014kkl,Finke:2015ona,VERITAS:2017gkr, AlvesBatista:2021sln}. 
	
	If magnetic fields have a primordial origin, then they tend to enhance baryon density perturbations after baryons decouple from photons in the early Universe \cite{wasserman97,Kim:1994zh,Subramanian:1997gi,gopalsethi03,Jedamzik:2013gua}. This is because a stochastically distributed magnetic field has a compressible component of the Lorentz force acting on the baryon fluid. When baryons are strongly coupled with the photon bath, the relativistic photon pressure counteracts any growth in baryon density perturbations. But on scales below the photon mean free path, the baryon fluid is decoupled from the photon bath and the compressible Lorentz force can lead to the growth of baryon inhomogeneities.
	
	Earlier works that studied the impact of primordial magnetic fields (PMFs) on the matter power spectrum focussed on length scales larger than the so called magnetic Jeans length and thereby focussed on structures bigger than $\sim 10^6\ M_{\odot}$  \cite{wasserman97,Kim:1994zh,gopalsethi03,Sethi:2004pe,Tashiro:2006,Schleicher:2008hc,sethi:2009,Shaw:2010ea,pandeysethi:2012,pandeysethi:2013,Chongchitnan:2013vpa,pandeysethi:2015,Minoda:2017iob,Sanati:2020,Katz:2021}. On scales smaller than the magnetic Jeans length, baryon perturbations tend to become non-linear prior to recombination \cite{Jedamzik:2013gua}. 
	However, the non-linearities in the baryon fluid eventually lead to turbulence at recombination and all signs of enhanced baryon density perturbations below the magnetic Jeans length are now erased \cite{Jedamzik:2018itu,Trivedi:2018ejz}.
	
	In this Letter, we show that even though the small-scale baryon density perturbations are erased, \pr{their growth history prior to recombination enhances the amplitude of dark matter (DM) density perturbations, solely through gravitational interactions}. Thus, searches for DM minihalos with masses smaller than $10^6 M_{\odot}$ can be used to probe PMFs.
	
	\noindent {\bf Magnetic fields in the photon drag regime.}  
	We shall focus on scales smaller than the photon mean free path and on times before recombination. Moreover, we consider scales larger than 0.01 pc, where the baryon fluid can be approximated as a perfect fluid and a perfect conductor prior to recombination. Consequently, the motion of the baryon fluid under the influence of a PMF is given by \cite{Subramanian:1997gi}
	\begin{align}\label{eq:vb}
		\frac{\partial \vec{v}_b}{\partial t}+\left(H+\alpha\right)\vec{v}_b+\frac{c_b^2}{a}\nabla \delta_b&=\frac{\vec{L}_B}{a}-\frac{\nabla\phi}{a},
	\end{align}
	where $a$ is the scale factor, \pr{$t$ is the cosmic time,} $\vec{v}_b=ad\vec{x}_b/dt$ is the physical baryon velocity, $\delta=(\rho(x)-\bar{\rho})/\bar{\rho}$ is the fluid density perturbation, $\phi$ is the metric perturbation following the convention in \cite{Ma:1995ey}, $\alpha=4\rho_{\gamma}/(3\rho_bl_{\gamma})$ parametrizes the photon drag force with $l_{\gamma}$ being the photon mean free path, $\rho_{\gamma}$ is the photon energy density, $c_b^2$ is the baryon sound speed and parametrizes the baryon thermal pressure, $H$ is the Hubble rate, and $\vec{L}_B=(\nabla\times\vec{B})\times\vec{B}/[4\pi a^4\rho_b]$ is the Lorentz force with $\vec{B}=a^2\vec{B}_{\rm phys}$ being the comoving magnetic field and $\vec{B}_{\rm phys}$ being the physical magnetic field. The evolution of $B$ is determined by,
	\begin{align}\label{eq:induction}
		\frac{\partial \vec{B}}{\partial t}&=\frac{1}{a}\nabla\times(\vec{v}_b\times \vec{B}).
	\end{align}
	We work in natural units where $\hbar=c=1$.
	
	In eq.~\eqref{eq:vb} we have ignored the contribution from convection, $(\vec{v}_b\cdot\nabla)\vec{v}_b/a$, because the viscous drag on baryons due to scattering with photons, $\alpha v_b$, is much larger than the convective term, \pr{or equivalently, the Reynolds number is much smaller than one} \cite{Banerjee:2004df}. One can verify this by using the fact that $\alpha$ is orders of magnitude larger than $H$ prior to recombination, $\alpha/H\sim 350 (a_{\rm rec}/a)^2$.
	
	In this section, we are interested in computing the evolution of the PMF power spectrum, $P_B(k,t)$. \pr{For simplicity, we shall focus on nonhelical PMFs, which have}
	\begin{align}\label{eq:def_M}
		\langle B_i(k) B^*_j(k')\rangle\!=\!\deltabar^3\!(k-k')\left(\delta_{ij}-\frac{k_ik_j}{k^2}\right)\frac{P_B(k)}{2},\!
	\end{align}
	where $B_i(k)=\int d^3x B_{i}(x) e^{ikx}$ and $\deltabar^3=(2\pi)^3\delta^3$. With the above convention, PMF energy density is simply $\rho_B\equiv\langle B^2\rangle/[8\pi]=\int d\Pi_q P_B(q)/[8\pi]$, where $d\Pi_q=d^3q/(2\pi)^3$.
	
	Taking the time derivative of eq.~\eqref{eq:def_M} and replacing $\partial B/\partial t$ using eq.~\eqref{eq:induction} would yield the evolution equations for $P_B$. We can make further analytical progress by noting that in eq.~\eqref{eq:vb}, $(\alpha+H) v_b\gg \partial v_b/\partial t\sim Hv_b$ before recombination. Then if the Lorentz force dominates over thermal pressure and gravity, we have $a\vec{v}_b\approx \vec{L}_B/(\alpha+H)$. Working in this large Lorentz force limit and neglecting non-Gaussianities in the distribution of $B$ (i.e. quasi-normal approximation), we obtain
	\begin{align}
		\frac{\partial P_B(k)}{\partial t}\!=&\frac{-4k^2P_B(k)}{3(\alpha+H)}\frac{\langle B^2\rangle}{4\pi\rho_{b}}.\!\label{eq:P_evolve}
	\end{align}
	\pr{Similar equation has been derived earlier for incompressible fluid in the viscous drag regime \cite{Campanelli:2013iaa} (for analytical treatments in the turbulence regime see for example Refs.~\cite{Campanelli:2007tc, Jedamzik:2010cy, Saveliev:2013uva}, and Ref.~\cite{Berger:2014wta} for analytical treatment of n-point correlations).}
	
	
	The solution to eq.~\eqref{eq:P_evolve} is given by
	\begin{align}\label{eq:PKi}
		P_B[k,k_D(t)]=P_B(k,t_i)e^{-k^2/k_D^2},
	\end{align}
	where $k_D$ represents the damping scale due to photon drag, and $P_B(k,t_i)$ is the power spectra prior to the photon drag regime, i.e. when the PMF coherence length scale was larger than the photon diffusion length, $l_{\gamma D}\approx 1/4\sqrt{l_{\gamma}/[aH]}$. 
	
	\pr{Prior to the photon drag regime, the baryon-photon plasma exhibits high Reynolds numbers. Consequently, PMFs induce turbulence in the baryon-photon plasma on small scales, with $P_B(k,t_i)\propto k^{-11/3}$ according to Kolmogorov cascade. On large scales, the power-law dependence of $P_B(k,t_i)$ is set by the physics of magnetogenesis. In this Letter, we consider PMFs with a Batchelor spectrum on large scales, $P_B(k,t_i)\propto k^2$, which is often encountered in magnetogenesis from phase transitions \cite{Durrer:2003ja,Vachaspati:2020blt,Subramanian:2015lua}.  We adopt the exact shape of $P_B(k,t_i)$ from Ref.~\cite{Mtchedlidze:2021bfy}, and we rescale it to achieve the desired initial PMF strength, $B_{I}\equiv \langle B^2(t_i)\rangle$, and the initial coherence length scale, $\xi_I\equiv B_I^{-2}\int d\Pi_q P_B(q,t_i)/q$. }
	
	\pr{In magnetogenesis from phase-transitions, $\xi_I$ is directly determined by $B_I$. In particular, if Alfv\'enic physics governs magneto-hydrodynamic (MHD) turbulence, then  $\xi_I\sim V_{AI}/[aH]_i$, with $V_{AI}=B_I/\sqrt{4\pi a^4(\rho_b+4\rho_{\gamma}/3)}$ \cite{Kahniashvili:2012uj, Banerjee:2004df}. Recently, it has been argued that a reconnection-controlled MHD turbulence leads to a better agreement with simulations \cite{PhysRevX.11.041005, Zhou:2022xhk}, leading to $\xi_I\sim 0.1 V_{AI}/[aH]_i$ for $B_I$ values of our interest \cite{Hosking:2022umv}. Due to uncertainties in turbulence theory, we keep $\xi_I$ as a free parameter. This also allows us to consider inflationary magnetogenesis scenarios, which can produce $\xi_I\gg V_{AI}/[aH]_i$.}
	
	\pr{In the photon drag regime, the damping scale can become the new coherence scale if $k_D^{-1}$ exceeds $\xi_I$}. Substituting  eq.~\eqref{eq:PKi} back in eq.~\eqref{eq:P_evolve}, we obtain
	\begin{align}\label{eq:kd_evolve}
		a\frac{\partial }{\partial a} \left(\frac{1}{k_D^2}\right)=\frac{4V_A^2/3}{a^2H(\alpha+H)},
	\end{align}
	where $V_A^2=\langle B^2\rangle /(4\pi a^4\rho_b)$ is the Alfv\'en speed. Given that $a\vec{v}_b\approx \vec{L}_B/(\alpha+H)$ and $L_B\sim k_DV_A^2$, eq.~\eqref{eq:kd_evolve} tells us that the damping scale is roughly the distance a baryon particle travels in a Hubble time, $ak_D^{-1}\sim v_b/H\sim k_DV_A^2/[aH(\alpha+H)]$. The above equation agrees with the free-streaming Alfv\'en damping scale derived for pre-recombination universe ($\alpha\gg H$) by Ref.~\cite{Jedamzik:1996wp}, as well as with the magnetic Jeans damping scale derived for post-recombination universe ($\alpha\ll H$) by Ref.~\cite{Kim:1994zh,Subramanian:1997gi}.
	
	We have neglected the effect of thermal pressure and gravity while deriving eq.~\eqref{eq:kd_evolve}. We find that even if thermal pressure dominates over the Lorentz force, the damping scale \pr{remains largely unchanged} \cite{supp} (see also \cite{Jedamzik:1996wp,Campanelli:2013iaa}). Furthermore, for PMFs that produce observable enhancement in the DM power spectrum, we find gravity to overcome the Lorentz force only after $k\gg k_D$. Consequently, we approximate the damping scale to be given by eq.~\eqref{eq:kd_evolve} in all regimes.
	
\begin{figure}
		\begin{center}
			\includegraphics[width=0.98\linewidth]{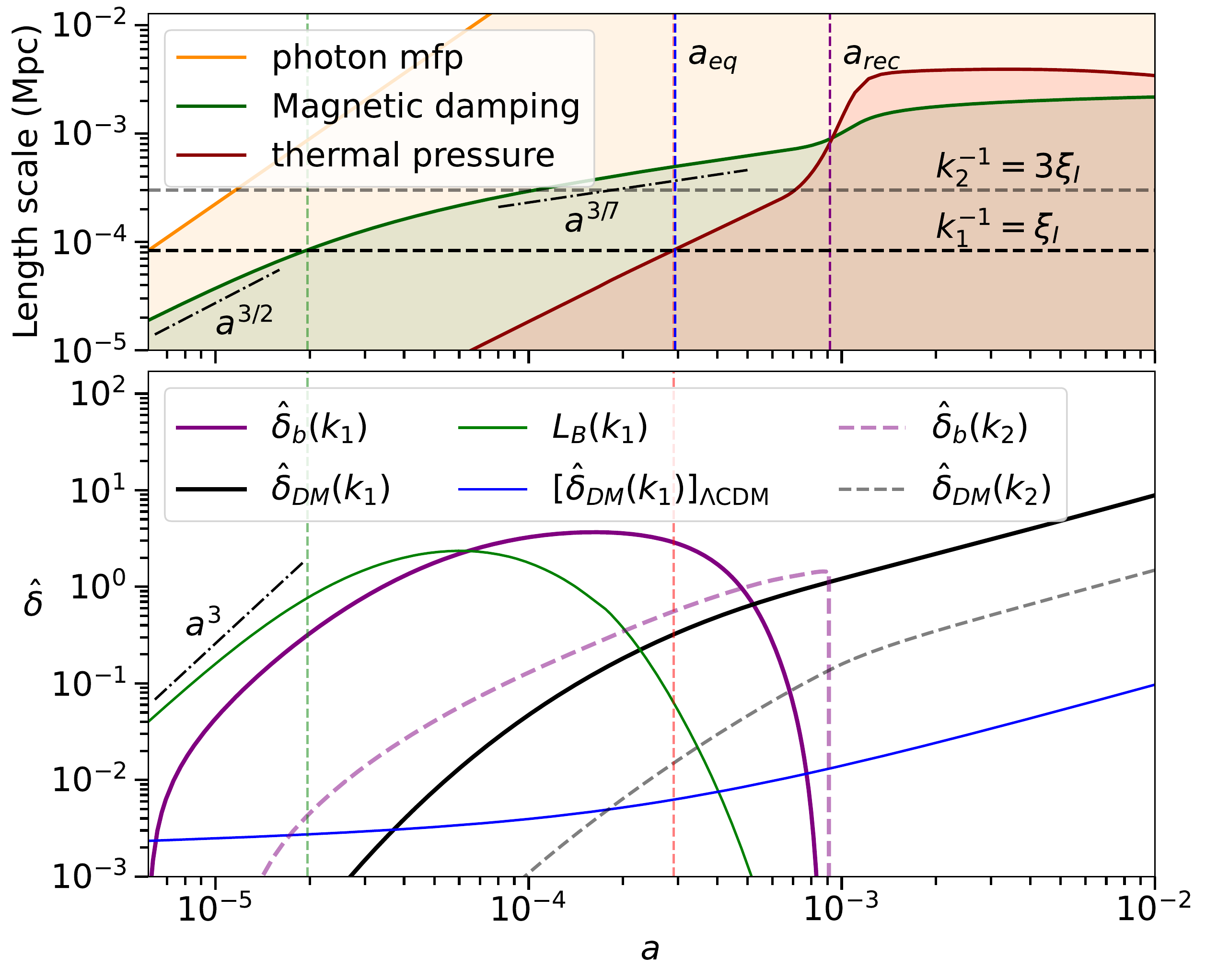}
			\caption{Here, $B_I=5$ nG and $\xi_I=10^{-4}$ Mpc. \textbf{Top:} Evolution of the photon mean free path (orange), the magnetic damping scale $k_D^{-1}$(green), and the thermal pressure damping scale $\lambda_{th}$(red). Horizontal dashed lines mark the comoving wave numbers corresponding to perturbations shown in the bottom panel. \textbf{Bottom:} Evolution of the square root of dimensionless power spectra, $\hat{\delta}=\sqrt{k^3P/(2\pi^2)}$. The green line shows $\hat{\delta}$ for $a^{-2}\nabla\cdot L_B/[H(\alpha+H)]$. The blue line shows $\hat{\delta}_{\rm DM}$ in a universe with no PMF, while other lines correspond to perturbations sourced by PMFs assuming trivial initial condition. Vertical green and red dashed lines mark the time when $k_1^{-1}$ crosses $k_D^{-1}$ and $\lambda_{th}$, respectively.
				\label{fig:delta_evolve}}
		\end{center}
	\end{figure}

	In the top panel of figure~\ref{fig:delta_evolve}, the green line shows the evolution of the damping scale, $k_D^{-1}$. We have solved the background cosmology parameters ($H$, $\alpha$, $\rho$, etc.) using CLASS \cite{CLASS}. One can see that $k_D^{-1}$ grows as $a^{3/2}$ when $k_D^{-1}\lesssim \xi_I$ and as $k_D^{-1}\propto a^{3/7}$ after $k_D^{-1}>\xi_I$. This transition occurs because when $k_D^{-1}<\xi_I$ the PMF is flux frozen (constant $\langle B^2\rangle$) but afterwards the PMF is damped as $\langle B^2\rangle\propto k_D^{5}\propto a^{-15/14}$. The sudden increase in $k_{D}^{-1}$ near recombination is due to the rapid decrease in photon drag, $\alpha$.

	\pr{After recombination, the rapid decrease in $\alpha$ causes the Reynolds number of the plasma to become larger than one \cite{Banerjee:2004df}.} Thus, the MHD fluid becomes turbulent and this turbulence is not captured in our formalism. Nevertheless, we extend eq.~\eqref{eq:kd_evolve} beyond recombination and find a logarithmic growth of $k_D^{-1}$, which is also expected from simulations \cite{Banerjee:2004df}. Consequently, our evaluation of present-day values of $B$ and $k_D$ should be accurate up to an order of magnitude.
	
	
	\noindent {\bf Impact on baryon density perturbations.} The continuity equation for $\delta_b$ in the linear limit is
	\begin{align}
		\frac{\partial \delta_b}{\partial t}+\frac{\nabla\cdot\vec{v}_b}{a}=0.
	\end{align}
	We find that the nonlinear term in the continuity equation, $\nabla\cdot(\delta_b\vec{v}_b)/a$, tends to enhance the power spectrum of $\delta_b$ by $\sim10\%$ when baryons are driven by the Lorentz force \cite{supp}. We found this by taking the ensemble average of the continuity equation convolved with $\delta_b(x')$ and using the linear solution, $\delta_b\propto \nabla\cdot v_b /[aH]=a^{-2}\nabla\cdot L_B/[H(\alpha+H)]$. As we are not interested in a precision calculation of the DM power spectrum, we have ignored the nonlinear term here.
	
	
	In the bottom panel of figure~\ref{fig:delta_evolve}, we show the evolution of $\hat{\delta}_b\equiv\sqrt{k^3P_b(k)/(2\pi^2)}$ for two different Fourier modes, where $P_b$ is the baryon power spectrum. 
	We start our computation when the photon mean free path exceeds the initial coherence length scale, $l_{\gamma}(t'_i)>\xi_I$. We set $\delta_b(t'_i)=v_b(t'_i)=0$, which is expected due to silk damping. Here $v_b$ is solved using eq.~\eqref{eq:vb}.
	
	The evolution of $\delta_b$ can roughly be classified into four regimes. 
	In the first regime, $\delta_b$ grows as $\sim a^3$ due to the Lorentz force. 
	Growth stops when the damping scale exceeds the length scale of the Fourier mode, $k_D^{-1}\gtrsim 3k^{-1}$, as the Lorentz force is damped out.
	
	Our evaluation of $\hat{\delta}_b$ for $k_D^{-1}\gtrsim k$ is not completely accurate. This is because we have approximated the evolution of $L_B(k,t)$ to be the same as $\sqrt{\langle L^2_B(k,t)\rangle}$, which is not true due to nonlinear processing between different Fourier modes of PMF.
	However, this approximation is acceptable for calculating $P_b(k,t)$, as $\delta_b\propto a^{-2}\nabla\cdot \vec{L}_B/(\alpha+H)$ when baryons are driven by $L_B$. Consequently, $P_b$ is directly determined by $\langle L^2_B(k,t)\rangle$. Furthermore, once $L_B$ is damped for $k\gg k_D$, $\delta_b$ \pr{becomes} insensitive to our approximation in $L_B(k)$. In the transition regime, we expect no more than $\mathcal{O}(1)$ correction to $P_b$.
	
	The second regime of $\delta_b$ evolution corresponds to saturation, where $v_b$ is quickly driven to zero by the photon drag and $\delta_b$ saturates to a constant. If baryon thermal pressure is ignored, $\delta_b$ asymptotes to $\sim\mathcal{O}(1)$ values for $k\gg k_D$. 
	The $\mathcal{O}(1)$ asymptote \pr{reflects} the fact that $k_D$ is determined by the distance traveled by a baryon particle in a Hubble time, $k_D^{-1}\sim v_b/(aH)$, which translates to $\delta_b\sim \mathcal{O}(1)$ as $\delta_b\sim kv_b/(aH)$.
	
	The third regime corresponds to the damping of $\delta_b$ by thermal pressure.
	The thermal pressure becomes important when it is \pr{comparable to} the photon drag force, $c_b^2k\delta_b\sim (H+\alpha)v_b$. 
	Using $\delta_b\sim kv_b/(aH)$, we find the thermal pressure damping scale to be
	\begin{align}
		\lambda_{th}=\frac{c_b}{aH}\sqrt{\frac{H}{\alpha+H}}.
	\end{align}
	We plot this scale as a red line in the top panel of figure~\ref{fig:delta_evolve}.
	
	Finally, the fourth regime corresponds to turbulence damping near recombination. At recombination, the photon drag force almost instantaneously becomes negligible and the convective terms in the baryon Euler equation (eq.~\eqref{eq:vb}) can no longer be ignored \cite{Banerjee:2004df,Trivedi:2018ejz,Jedamzik:2018itu}. We model the impact of turbulence on $\delta_b$ by setting $\delta_b=0$ at recombination for all scales inside the magnetic damping scale.
	
	\noindent {\bf Impact on DM perturbations.} The rapid growth in $\delta_b$ gravitationally induces growth in DM perturbations, $\delta_{\rm DM}$. We find that $\delta_{\rm DM}$ largely remains in the perturbative regime before recombination even if $\delta_b$ reaches nonlinear values. This is because the gravitational influence of baryons is suppressed by a factor of $\rho_b/\rho_{\rm tot}$. Consequently, the growth in $\delta_{\rm DM}$ is well captured by the linear theory \cite{Hu:1995en,Meszaros:1974tb},
	\begin{multline}
		\frac{\partial^2 \delta_{\rm DM}}{\partial y^2}+\frac{2+3y}{2y(y+1)}\frac{\partial \delta_{\rm DM}}{\partial y}-\frac{3}{2y(y+1)}\frac{\Omega_{\rm DM}}{\Omega_m}\delta_{\rm DM}\\=\frac{3}{2y(y+1)}\frac{\Omega_{b}}{\Omega_m}\delta_{b}.
	\end{multline}
	Here $\Omega$ corresponds to the present-day fraction of species, with $\Omega_bh^2=0.022$, $\Omega_{\rm DM}h^2=0.12$, $\Omega_m=\Omega_b+\Omega_{\rm DM}$, and $y=a/a_{eq}$ with $a_{eq}$ being the scale factor at matter radiation equality \cite{Aghanim:2018eyx}.
	
	Since $\delta_{\rm DM}$ follows an ordinary differential equation, its solution can simply be written as a linear combination of the homogenous solution provided by the initial condition and the inhomogeneous solution indirectly sourced by PMF, $\delta_{\rm DM}=(\delta_{\rm DM})_{\Lambda CDM}+(\delta_{\rm DM})_{B}$. As the PMF distribution is uncorrelated with the curvature perturbations determining initial conditions \cite{Aghanim:2018eyx}, the two solutions of $\delta_{\rm DM}$ are uncorrelated as well \cite{Kunze:2022mlr}.
	
	
	
	\begin{figure}
		\begin{center}
			\includegraphics[width=0.98\linewidth]{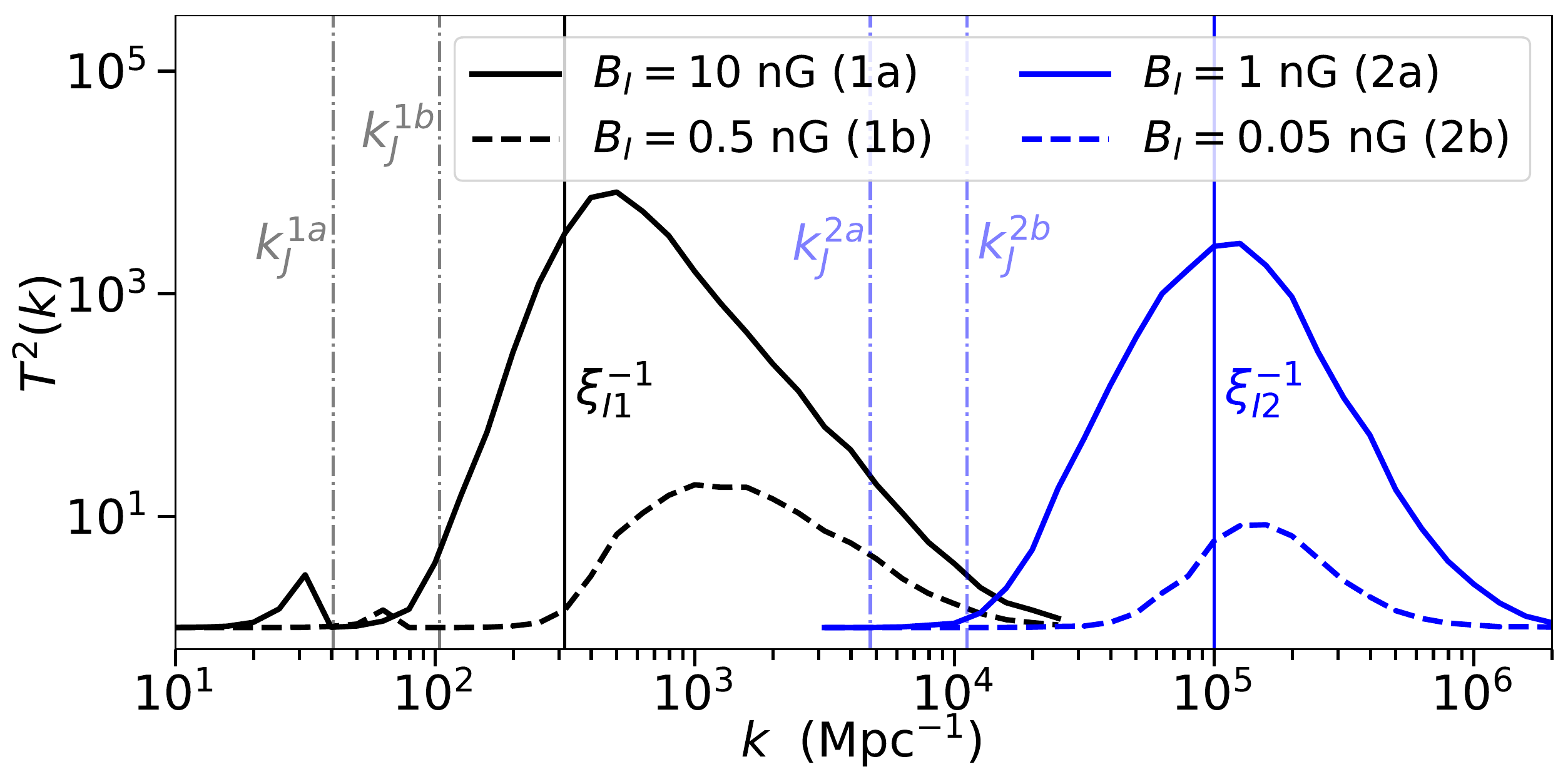}
			\caption{The dark matter transfer function for PMF configurations sampling the edges of the parameter space shown in figure~\ref{fig:param_space}. \pr{The dot-dashed lines mark the magnetic Jeans scale, $\sim k_J=k_D(2a_{rec})$.}
				\label{fig:Tk}}
		\end{center}
	\end{figure}
	
	In figure~\ref{fig:Tk}, we show the DM transfer function, $T^2(k)=P_{\rm DM}(k)/[P_{\rm DM}(k)]_{\Lambda CDM}$, where $P_{\rm DM}$ is the DM power spectrum, $\deltabar^3(k-k')P_{\rm DM}(k)=\langle\delta_{\rm DM}(k) \delta_{\rm DM}(k')\rangle_B+\langle\delta_{\rm DM}(k) \delta_{\rm DM}(k')\rangle_{\Lambda CDM}$. The transfer function is quite narrowly peaked with the peak typically occurring at $k_{\rm pk}\sim \xi_I^{-1}$. 
	For $k<k_{\rm pk}$, thermal pressure or turbulence at recombination suppresses $\delta_b$ before it attains its maximum value due to the Lorentz force. The value of $k_{\rm pk}$ for the black dashed line is shifted to $k_{\rm pk}>\xi_I^{-1}$, because $\delta_b(k=\xi_I^{-1})$ is prematurely suppressed by turbulence at recombination. The value of $T(k)$ for $k>\xi_I$ is suppressed because the initial PMF is itself suppressed.
	
	\begin{figure}
		\begin{center}
			\includegraphics[width=\linewidth]{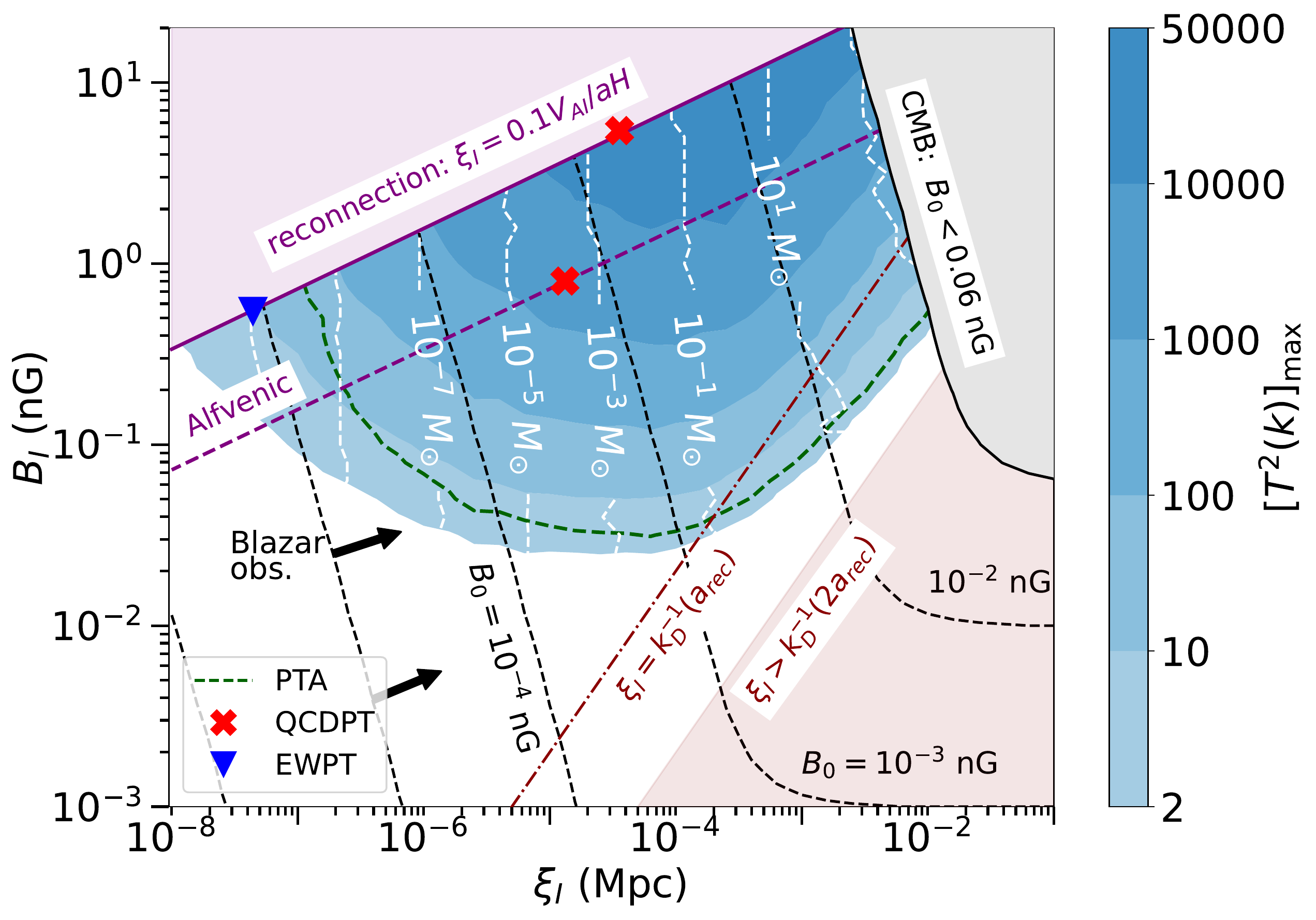}
			\caption{Colored contours of the maximum value of dark matter transfer function as a function of initial comoving magnetic field strength, $B_{I}$, and initial coherence length scale, $\xi_I$. Here initial refers to the time \pr{before recombination, when the photon diffusion length exceeds $\xi_I$}. White contours show the mass scales at which the transfer function reaches its peak. Black dashed lines show the contours of present-day magnetic field strength, $B_0$. The region inside the green-dashed line produces dark matter minihalos potentially within the observable reach of PTAs. Solid and dashed purple lines denote the relation between $\xi_I$ and $B_I$ depending on whether MHD turbulence is determined by reconnection or Alfv\'enic physics, respectively. The red crosses (blue triangle) mark the maximum values of $B_I$ for PMFs produced during QCD (electroweak) phase transition. \label{fig:param_space}}
		\end{center}
	\end{figure}
	
	\noindent {\bf Connecting PMFs with DM minihalos.}
	In figure~\ref{fig:param_space} we vary over different initial PMF configurations and plot the contours showing the maximum enhancement to the DM transfer function, $[T^2(k)]_{\rm max}$.
	A large value of $[T^2(k)]_{\rm max}$ implies that the halos corresponding to the peak scale, $k_{\rm pk}^{-1}$, collapse earlier than in the standard cosmology and consequently have much larger central densities \cite{Bullock:1999he,Wechsler:2001cs,Delos:2019mxl}. The mass of these first formed minihalos is approximately given by the DM mass enclosed in a radius of $ak_{\rm pk}^{-1}$, i.e. $M_h=4\pi\rho_{\rm DM}(a=1)k_{\rm pk}^{-3}/3$.
	
	
	The value of $[T^2(k)]_{\rm max}$ is suppressed for small $B_I$ and for both large and small $\xi_I$. When $B_{I}<0.05$ nG, the thermal presssure inhibits the growth in $\delta_b(k_{\rm pk})$ before it attains its $\mathcal{O}(1)$ saturation value, which then limits the growth in $\delta_{\rm DM}$. For $\xi_I<10^{-4}$ Mpc, $\delta_b(k_{\rm pk})$ is suppressed by thermal pressure before $a_{eq}$. Consequently, the gravitational influence of baryons on DM is suppressed by a factor of $\sim a/a_{eq}$. Finally, for $\xi_I\gtrsim k_D^{-1}(a_{\rm rec})$, turbulence near recombination suppresses $\delta_b$ before it can attain $\mathcal{O}(1)$ values. Thus, we have restricted our analysis to $\xi_I<k_D^{-1}(2a_{\rm rec})$.
	
	
	
	Additionally, $\xi_I$ has a natural lower bound for a given value of $B_I$. In particular, prior to the photon drag regime, PMFs induce turbulent motion in the plasma and are damped by Kolmogorov cascade below the turbulence scale. Thus, $\xi_I$ has to be larger than the purple lines in figure~\ref{fig:param_space}, depending on the physics controlling MHD turbulence.
	
	
	
	
	If PMFs are generated after inflation, then turbulence typically determines the coherence length scale of PMFs and $B_I$ and $\xi_I$ would lie on one of the purple lines. For instance, considering magnetogenesis at QCD phase transition, the largest value of $B$ is obtained by assuming $\rho_B=\rho_{SM}$ and $\xi=(aH)^{-1}$ at $T_{QCD}=150$ MeV. The subsequent turbulent evolution increases $\xi$ while conserving $B^4\xi^5$ ($B^2\xi^5$) for reconnection-controlled (Alf\'venic) turbulence \cite{PhysRevX.11.041005}.
	
	PMFs generated after inflation can explain the absence of GeV gamma-ray halos around TeV blazars, if their present day strength, $B_0=\langle B^2(a=1)\rangle$, is larger than $\sim 10^{-5}$ nG \cite{Taylor:2011bn}. One can see from figure~\ref{fig:param_space} that PMFs generated after electroweak phase transition (EWPT) can explain Blazar observations \cite{Hosking:2022umv} and also enhance $T^2(k)$.
	
	An enhancement in $T^2(k)$ implies an enhanced abundance of minihalos today.
	The presence of such minihalos in the Milkyway can potentially be observed by astrometric microlensing searches \cite{Erickcek:2010fc,Li:2012qha,VanTilburg:2018ykj}, caustic microlensing \cite{Oguri:2017ock,Diego:2017drh}, or by pulsar timing arrays (PTAs) \cite{Siegel:2007fz,Clark:2015sha,Ramani:2020hdo,Lee:2020wfn}. 
	For instance, the region inside the green-dashed line has $P_{\rm DM}(k)$ which is potentially in the observable window of future PTA measurements \cite{Lee:2020wfn}. Note that this is an \pr{optimistic} estimate \cite{Lee:2021zqw}, and a more detailed analysis is required to draw any definitive conclusion on observability.
	
	
	The current strongest constraint restricts $B_0<0.06$ nG \pr{ at $95\%$ confidence level for PMFs with a Batchelor spectrum} \cite{Paoletti:2022gsn}. Further stringent constraint of $B_0<0.01$ nG can be applied if we take into account that non-linear values of $\delta_b$ can alter recombination \cite{Jedamzik:2018itu}. Recently, this fact has instead been utilized to help resolve the Hubble tension \cite{Jedamzik:2020krr}. Depending on the exact values of $B_I$ and $\xi_I$ that resolve the Hubble tension, it is possible that an observable abundance of DM minihalos is also produced. 
	
	\noindent {\bf Discussion.} 
	We have shown for the first time that primordial magnetic fields (PMFs) can enhance the present-day matter power spectrum on scales below the magnetic Jeans length. This implies a larger abundance of DM minihalos today.
	
	The abundance of DM minihalos is determined not by the present-day strength of the PMF, but rather by its strength in the early Universe when baryon perturbations on the PMF coherence scale decoupled from photons. As PMFs had underwent less dissipation at that time, they had a larger potential to overcome baryon thermal pressure and generate baryon inhomogeneities. DM acts as a ``memory foam" and preserves the growth history of these inhomogeneities. As a result, DM minihalos can probe PMFs with present-day strengths much below those probed by cosmic microwave background (CMB) and Faraday rotation.
	
	Considering nonhelical PMFs that are generated after inflation with a Batchelor spectrum, we estimate that PMFs that can explain blazar observations should also produce DM minihalos heavier than $10^{-10}\ M_{\odot}$. These minihalos may be detectable through future PTA measurements. 

	\pr{Our results are based on analytical calculations, with the primary assumption that the distribution of PMFs remains Gaussian. Our analytical approach agrees with previous results in the literature, suggesting that non-Gaussian effects have a limited role in the photon drag regime. Consequently, even with dedicated MHD simulations, we anticipate that the qualitative relationship between the DM power spectrum and present-day PMF strengths will remain largely unchanged. However, there may be some variation in the quantitative relationship, possibly within an order of magnitude.}
	
	\pr{Even if non-gaussianities in the PMF distribution are significant, we expect the DM minihalos to still be produced below the magnetic jeans scale. This is simply because the magnetic jeans scale is set by the scale where baryon perturbations become large enough to backreact onto the magnetic fields. These large baryon perturbations gravitationally enhance DM perturbations. Subsequently, even after baryon perturbations are damped by PMFs, the gravitational potential from DM is sufficient to sustain the growth in DM perturbations and cause their eventual collapse into minihalos.}
	
	Thus, a future detection of DM minihalos should serve as a smoking gun signal for a primordial origin of cosmic magnetic fields. It is rather ironic how we can utilize the invisible component of our Universe to search for a component of the visible sector.
	
	\medskip
	\begin{acknowledgments}
		\noindent {\it Acknowledgments} The author thanks Karsten Jedamzik for clarification about initial magnetic field configurations and cross-checking some calculations, Kandaswamy Subramanian for initial guidance on relevant literature, Takeshi Kobayashi for several helpful discussions, Adrienne Erickcek for clarification on minihalos and helpful discussions, Andrea Mitridate for clarification about PTA sensitivities, and Andrey Saveliev for clarifications about his work and highlighting different conventions used for PMF power spectra. The author is also thankful to Karsten Jedamzik, Takeshi Kobayashi, and Jan Schutte-Engel for helpful feedback on the manuscript.
	\end{acknowledgments}
	
	\bibliography{references}
	
	\newpage
	\onecolumngrid

	\appendix
	
	\section{Magnetic damping scale for incompressible fluid}
	In this section we derive the evolution of non-helical magnetic field power spectrum for scenarios where the baryon thermal pressure dominates over the Lorentz force in the photon drag regime.
	
	In the main Letter we showed that the baryon velocity is determined by the balance between the Lorentz force and the photon drag force, $a\vec{v}_b\approx \vec{L}_B/(\alpha+H)$, when thermal pressure is negligible.
	In contrast, in the strong thermal pressure limit, compressible motion in the baryons is suppressed, $\nabla\cdot v_b\rightarrow 0$. However, the Lorentz force can \textbf{still} induce divergence free motion in the baryon fluid without hindrance.
	Consequently, one can write the Fourier transformed baryon velocity vector as
	\begin{align}\label{eq:vb_incomp}
		v_{b,j}(q)=\left(\delta_{lj}-\frac{q_lq_j}{q^2}\right)\frac{L_{B,l}(q)}{a(\alpha+H)}=\hat{p}_{lj}(q)\frac{L_{B,l}(q)}{a(\alpha+H)},
	\end{align}
	where $\vec{L}_B(q)$ is the Fourier transform of the Lorentz force, $\vec{L}_B(x)=(\nabla\times\vec{B})\times\vec{B}/[4\pi a^4\rho_b]$.
	
	The motion in the baryon fluid feedbacks onto the magnetic field via the induction equation, eq.~\eqref{eq:induction}. Taking the Fourier transform of eq.~\eqref{eq:induction} and replacing $v_b$ using eq.~\eqref{eq:vb_incomp}, we obtain
	\begin{multline}\label{eq:temp}
		\frac{\partial B_i(k)}{\partial t}=\frac{-k_m }{(4\pi\rho_{b})(\alpha+H)}[\delta_{ml}\delta_{ji}-\delta_{mj}\delta_{li}][\delta_{ab}\delta_{nc}-\delta_{bc}\delta_{na}]\int d\Pi_{q_1} d\Pi_{q_2}\hat{p}_{jn}(q_1)q_{2a}B_b(q_1-q_2)B_c(q_2) B_l(k-q_1),
	\end{multline}
	where $d\Pi_q=d^3q/(2\pi)^3$.
	
	To obtain the evolution of the power spectrum, we use the fact that 
	\begin{align}
		\deltabar^3(k-k')\frac{\partial P_B(k)}{\partial t}=\frac{1}{2} \left\langle\frac{\partial B_m(k)}{\partial t}B_m^{*}(k')\right\rangle.
	\end{align}
	Then replacing the derivative of $B$ using eq.~\eqref{eq:temp}, using $\langle B_i(k)B_j^*(k')\rangle\equiv \deltabar^3(k-k')M_{ij}(k')$, and assuming magnetic fields to be Gaussian distributed, we obtain
	\begin{multline}
		\frac{\partial P_B(k)}{\partial t}
		=\frac{-\deltabar^3(k-k')}{2(4\pi\rho_{b})(\alpha+H)}k_i[\delta_{il}\delta_{jm}-\delta_{ij}\delta_{lm}][\delta_{ab}\delta_{nc}-\delta_{bc}\delta_{na}]\\
		\times\int d\Pi_q\hat{p}_{jn}(k-q) [k_{a}M_{lb}(q)M_{cm}(k)-q_{a}M_{lc}(q)M_{bm}(k)].
	\end{multline}
	Next, considering non-helical fields, we replace $M$ using eq.~\eqref{eq:def_M} and set $h=0$. Doing so yields
	\begin{align}
		\frac{\partial P_B(k)}{\partial t}=\frac{-P_B(k)}{(4\pi\rho_{b})(\alpha+H)}\int d\Pi_q \left[-\frac{2}{3}(k^2+3q^2)+\frac{(k^4+k^2q^2+q^4)}{kq}\ln\left(\frac{k^2+q^2+kq}{k^2+q^2-kq}\right)\right]P_B(q).
	\end{align}
	
	To obtain a simple expression we can look at the asymptotic limit, $k\ll k_D$ or $k\gg k_D$. In these limits, the integral is dominated by the phase space where $q\sim k_D\gg k$ or $q\sim k_D\ll k$. Both limits yield
	\begin{align}\label{eq:kd_th}
		\frac{\partial P_B(k)}{\partial t}\approx-\frac{4k^2V_A^2}{3a^2(\alpha+H)}P(k),
	\end{align} 
	where $V_A^2=\langle B^2\rangle/[4\pi a^4\rho_b]$ is the Alfv\'en speed.
	
	Thus, we can see that the magnetic power spectrum evolution is well approximated by $P(k,t)=P(k,t_i)e^{-k^2/k_D^2}$ for $k$ values well separated from $k_D$. The evolution of the damping scale, $k_D^{-1}$, can be obtained by plugging this solution back in eq.~\eqref{eq:kd_th}. Doing so surprisingly yields the exact same equation as for compressible fluid, eq.~\eqref{eq:kd_evolve}.
	
	While deriving the power spectrum evolution here and in the main Letter, we considered that $v_b\propto L_B$ on all scales. However, for $k>k_D$, the magnetic fields are being exponentially damped and for large enough $k$, the baryon flow should stop following the Lorentz force. However, as long as the scale where baryons stop following the Lorentz force is much smaller than $k_D^{-1}$ we can safely assume $v_b\propto L_B$ to always be valid for the purpose of evaluating $P_B(k,t)$. In particular, one can see that $L_B$ is roughly suppressed as $\sim k^2/k_D^2$ for $k>k_D$ while $v_b$ can at best decrease with a dimensionless rate of $(H+\alpha)/H$. Consequently, $v_b$ follows the Lorentz force until $k\gtrsim \sqrt{\alpha/H}k_D$. Prior to recombination, $\alpha$ is atleast 300 times larger than $H$, so baryons follow the Lorentz force deep into the damping regime.
	
	\section{Non-linear term in the baryon continuity equation}
	In this section we show that the non-linear term in the baryon continuity equation tends to enhance the baryon power spectrum when baryons are driven by magnetic fields.
	
	We begin with the baryon continuity equation,
	\begin{align}\label{eq:deltab_cont}
		\frac{\partial \delta_b}{\partial t}+\frac{\nabla\cdot\vec{v}_b}{a}+\frac{\nabla\cdot(\vec{v}_b\delta_b)}{a}&=0.
	\end{align}
	Taking the Fourier transform of the above equation and then ensemble averaging with $\delta_b^*(k')$, we obtain
	\begin{align}
		\langle\frac{\partial \delta_b(k)}{\partial t}\delta_b^*(k')\rangle+\frac{\langle \vec{k}\cdot\vec{v}_b(k)\delta_b^*(k')\rangle}{a}+\deltabar^3(k-k')\Xi(k)&=0,
	\end{align}
	where $\deltabar^3(k)=(2\pi)^3\delta(k)$ and 
	\begin{align}\label{eq:def_xi}
		\deltabar^3(k-k')\Xi(k)\equiv -\frac{i}{a}\int\frac{d^3q}{(2\pi)^3}\langle (\vec{k}\cdot \vec{v}_b(q))\delta_b(k-q)\delta_b^*(k')\rangle.
	\end{align}
	Here $\Xi(k)$ represents the contribution from the non-linear term in eq.~\eqref{eq:deltab_cont} to the evolution of the baryon density power spectrum.
	
	We are primarily interested in the regime where baryon perturbations become non-linear, i.e. where the Lorentz force dominates over the baryon thermal pressure and gravity. Consequently, $v_b$ is determined by the balance between the Lorentz force and the photon drag force, $a\vec{v}_b\approx \vec{L}_B/(\alpha+H)$. Furthermore, assuming that $\delta_b$ is largely determined by the linear continuity equation, eq.~\eqref{eq:deltab_cont}, we have $\delta_b=-C \nabla\cdot v_b/[aH]=-C\nabla\cdot\vec{L}_B/[a^2H(\alpha+H)]$, where $C$ is some constant of motion. As both $v_b$ and $\delta_b$ are now expressed in terms of the Lorentz force, we can simplify eq.~\eqref{eq:def_xi} to obtain $\Xi$ in terms of the magnetic field power spectrum,
	\begin{multline}
		\Xi(k) 
		= \frac{-4C^2[\delta_{ml}k_{j}-\delta_{mj}k_{l}](k_{r}k_{s}-k^2\delta_{rs}/2)}{(4\pi a^4\rho_b)^3 a^{6}(H+\alpha)^3H^2}\int d\Pi_qd\Pi_{q_1} [(k-q-q_1)_{u}(k-q-q_1)_{v}-(k-q-q_1)^2\delta_{uv}/2]\\ \times(q_1-q)_mM_{jr}(q_1)M_{lu}(q)M_{vs}(k-q_1),
	\end{multline}
	where $M$ is defined through $\langle B_i(k)B_j^*(k')\rangle\equiv \deltabar^3(k-k')M_{ij}(k')$ and $d\Pi_q=d^3q/(2\pi)^3$. While obtaining the above expression, we assumed the magnetic fields to be Gaussian distributed. In what follows, we shall show that $\Xi$ is largely negative and hence leads to an enhancement of the baryon power spectrum.
	
	We can further simplify the above integral by considering non-helical magnetic fields with $M$ given by eq.~\eqref{eq:def_M} after setting $h(k)=0$. Then after contracting over all indices and integrating over $q$, we obtain
	\begin{multline}
		\Xi(k)= \frac{C^2}{(4\pi a^4\rho_b)^2 a^{6}(H+\alpha)^3H^2} \frac{1}{8}\frac{2V_A^2}{3}k^3\int d\Pi_{q_1}
		P_B(q_1)\frac{P_B(k-q_1)}{(\vec{k}-\vec{q}_1)^2}\\
		\bigg(kq_1\bigg[q_1^3(1-3x^2+2x^4)-kq_1^2x(1-5x^2+4x^4)+k^2q_1(1-5x^2+4x^4)+k^3x(1-x^2)\bigg]\\+k_{*2}^2k_D^2\bigg[2q_1^3x^3-2kq_1^2x^2(x^2+2)+3k^2q_1x(1+x^2)-k^3(1+x^2)\bigg]\bigg).
	\end{multline}
	where $x=(\vec{k}\cdot \vec{q}_1)/kq_1$, $V_A^2=\int d\Pi_q P_B(q)/[4\pi a^4\rho_b]$, and $k_{*n}$ represents the moments of the magnetic power spectrum,
	\begin{align}
		k_{*n}^n\equiv \frac{\int d\Pi_q q^nP_B(q)}{\int d\Pi_q P_B(q)}.
	\end{align}
	
	The integral in $\Xi(k)$ is similar to the integral in the power spectrum of the Lorentz force \cite{Planck:2015zrl},
	\begin{multline}\label{eq:LB_power}
		\langle \nabla\cdot \vec{L}_B(k) \nabla\cdot \vec{L}_B(k')\rangle=\deltabar^3(k-k')P_{\nabla\cdot \vec{L}_B}=\deltabar^3(k-k')\frac{k^4}{8(4\pi a^4\rho_b)^2}\int \frac{d^3q}{(2\pi)^3} \frac{P_B(q)P_B(k-q)}{(k-q)^2}\bigg[k^2(1+x^2)-4kqx^3\\+2q^2(1-2x^2+2x^4)\bigg].
	\end{multline}
	We have checked that our Lorentz force power spectrum matches to that used in literature \cite{Planck:2015zrl}.
	\begin{figure}
		\centering
		\includegraphics[width=0.5\textwidth]{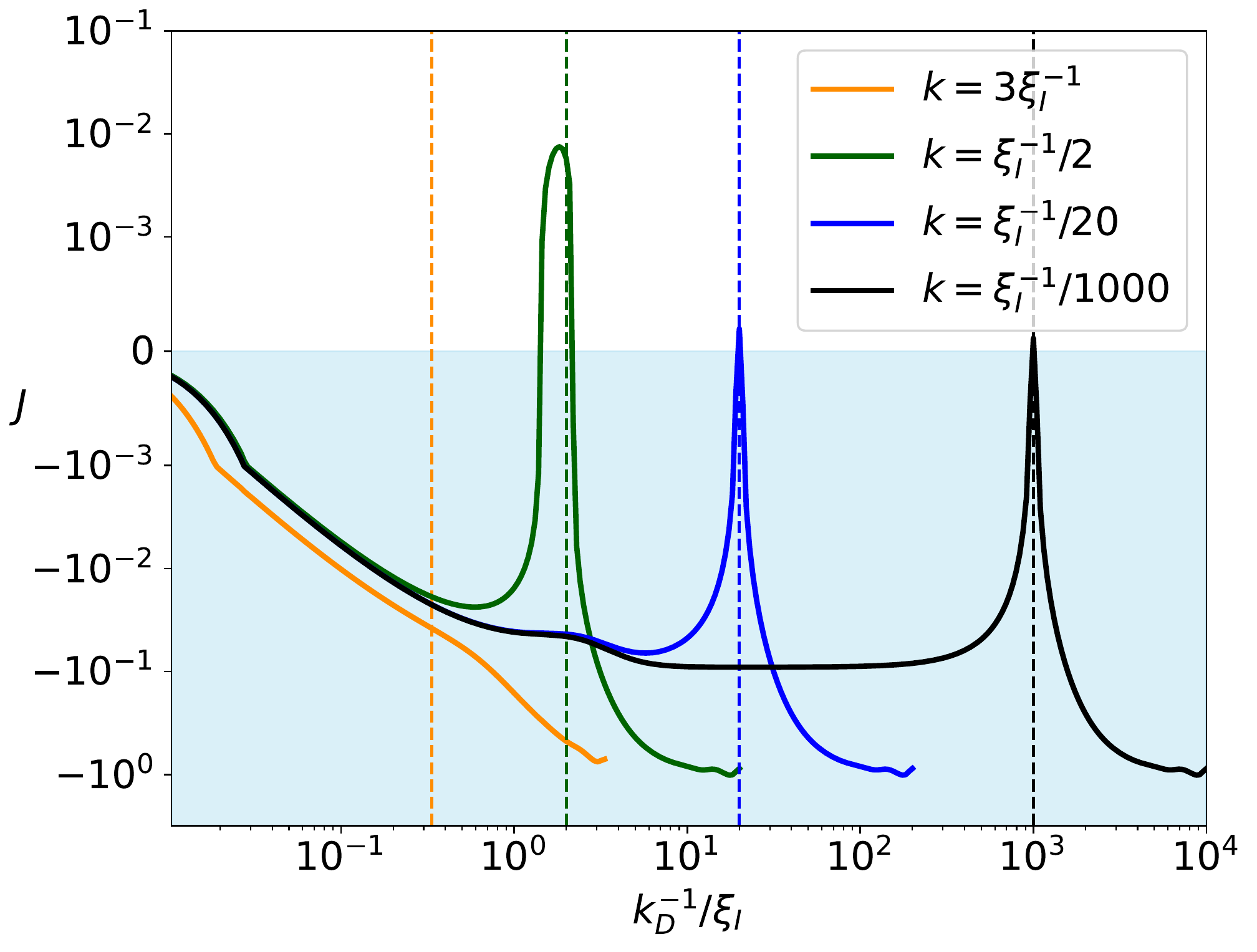}
		\caption{Figure shows $J$, which is a dimensionless quantity that parametrizes the impact of the non-linear term in the baryon continuity equation, as a function of the damping scale, as a function of the damping scale, $k_D^{-1}$, for different values of $k$. The vertical dashed lines mark the points where $k_D=k$. We can see that $J$ is almost always negative, except for small region near $k=k_D$. The negative values of $J$ imply that the non-linear term enhances the linear baryon power spectrum.} \label{fig:J_evolve}
	\end{figure}
	
	Consequently, we rewrite $\Xi$ in terms of $P_{\nabla\cdot \vec{L}_B}$ and use the fact that we have assumed $\delta_b=-C\nabla\cdot\vec{L}_B/[a^2H(\alpha+H)]$, to obtain
	\begin{align}
		\Xi(k)=\frac{C^2P_{\nabla\cdot \vec{L}_B}}{a^{6}(H+\alpha)^3H^2}\frac{2k_D^2V_A^2}{3}J(k,k_D)=\frac{2P_{b}(k)}{3}\frac{k_D^2V_A^2}{a^{2}(H+\alpha)}J(k,k_D),
	\end{align}
	where $\deltabar^3(k-k')P_b(k)=\langle \delta_b(k)\delta_b^*(k')\rangle$, and $J$ is a dimensionless quantity that parametrizes the ratio of the integrals in $\Xi$ and $P_{\nabla\cdot \vec{L}_B}$.
	
	Since $\Xi$ is proportional to $P_b$, we can effectively rewrite the continuity equation for baryon as
	\begin{align}\label{eq:deltab_cont2}
		\frac{\partial \delta_b}{\partial t}+\frac{\nabla\cdot\vec{L}_B}{a(\alpha+H)}+\frac{2\delta_b}{3}\frac{k_D^2V_A^2}{a^{2}(H+\alpha)}J(k,k_D)&=0,
	\end{align}
	where we have used $a\vec{v}_b\approx \vec{L}_B/(\alpha+H)$.
	The above equation would yield the same equation for $P_b$ as the original continuity equation when baryons are driven by the Lorentz force.
	
	The above equation also allows us to confirm our original assumption of $\delta_b=-C\nabla\cdot\vec{L}_B/[a^2H(\alpha+H)]$. To do so we simply substitute $\delta_b=-C\nabla\cdot\vec{L}_B/[a^2H(\alpha+H)]\propto a^{m}$ in eq.~\eqref{eq:deltab_cont2} to obtain,
	\begin{align}
		C=\frac{1}{m+\left[ \frac{2}{3}\frac{k_D^2V_A^2}{a^{2}(H+\alpha)}J(k,k_D)\right]}.
	\end{align}
	Note that $\frac{k_D^2V_A^2}{a^{2}(H+\alpha)}\sim \mathcal{O}(1)$ according to the definition of $k_D$ (eq.~\eqref{eq:kd_evolve}). Consequently, the non-linear term in the continuity equation simply provides a correction of order $J/m$ to the linear solution. In a radiation-dominated universe, we have $m=3$ prior to the damping of the magnetic fields.
	
	In figure~\ref{fig:J_evolve}, we show $J$ as a function of $k_D$ for different values of $k$. Note that $J$ is typically of order $-0.1$ for $k<k_D$. For $k\sim k_D$ we see $J$ to jump to a small positive value and then it again becomes negative. The large negative values of $J$ for $k\gg k_D$ have no physical significance because in that limit the magnetic fields and $v_b$ are exponentially damped. Consequently, our calculation of $\Xi$, which is based on the assumption of $\delta_b\propto \vec{v}_b/a$, is no longer applicable. We instead expect $\Xi$ to go to zero for $k\gg k_D$ because $\delta_b$ is no longer correlated with $\vec{v}_b\approx \vec{L}_B/(\alpha+H)$.
	
	
	Hence, when the growth in $\delta_b$ is sourced by the magnetic fields, we expect the non-linear terms in the baryon continuity equation to provide a small enhancement to the baryon power spectrum obtained from linear analysis.

\end{document}